\documentclass[10pt,twocolumn,letterpaper]{article}

\usepackage{cvpr}
\usepackage{times}
\usepackage{epsfig}
\usepackage{graphicx}
\usepackage{amsmath}
\usepackage{amssymb}
\usepackage[percent]{overpic}
\usepackage{boldline}
\usepackage{multirow,enumitem}
\usepackage{booktabs}

\newcolumntype{C}[1]{>{\centering}m{#1}}

\newcommand{\SKIP}[1]{}

\usepackage[pagebackref=true,breaklinks=true,letterpaper=true,colorlinks,bookmarks=false]{hyperref}

\cvprfinalcopy % *** Uncomment this line for the final submission

\def\cvprPaperID{20} % *** Enter the CVPR Paper ID here

\ifcvprfinal\fi
\begin{document}
	\title{ImagePairs: Realistic Super Resolution Dataset via Beam Splitter Camera Rig }
	
	\author{Hamid Reza Vaezi Joze \quad Ilya Zharkov \quad Karlton Powell \quad Carl Ringler  \quad Luming Liang \\ Andy Roulston \quad  Moshe Lutz \quad Vivek Pradeep   
		\\
		Microsoft\\
		%One Microsoft Way, Redmond WA, USA\\
		%{\tt\small \{hava,zharkov,kapowell, carlr, lulian, v-androu,moshelu,vpradeep\}@microsoft.com} }
	    {\tt\small\url{www.microsoft.com/en-us/research/project/imagepairs}} }
	\maketitle
	\thispagestyle{empty}
	
	%%%%%%%%% ABSTRACT
	\begin{abstract}
		Super Resolution is the problem of recovering a high-resolution image from a single or multiple low-resolution images of the same scene. It is an ill-posed problem since high frequency visual details of the scene are completely lost in low-resolution images. To overcome this, many machine learning approaches have been proposed aiming at training a model to recover the lost details in the new scenes. Such approaches include the recent successful effort in utilizing deep learning techniques to solve super resolution problem. As proven, data itself plays a significant role in the machine learning process especially deep learning approaches which are data hungry. Therefore, to solve the problem, the process of gathering data and its formation could be equally as vital as the machine learning technique used. Herein, we are proposing a new data acquisition technique for gathering real image data set which could be used as an input for super resolution, noise cancellation and quality enhancement techniques. We use a beam-splitter to capture the same scene by a low resolution camera and a high resolution camera. Since we also release the raw images, this large-scale dataset could be used for other tasks such as ISP generation.  Unlike current small-scale dataset used for these tasks, our proposed dataset includes 11,421 pairs of low-resolution high-resolution images of diverse scenes. To our knowledge this is the most complete dataset for super resolution, ISP and image quality enhancement. The benchmarking result shows how the new dataset can be successfully used to significantly improve the quality of real-world image super resolution.
	\end{abstract}
	
	%%%%%%%%% BODY TEXT
	\section{Introduction}
	Super Resolution (SR) is the problem of recovering high-resolution (HR) image from a single or multiple low-resolution (LR) images of the same scene. In this paper we are focusing on single-image SR which uses a single LR image as input. It is an ill-posed problem as the high frequency visual details of the scene are lost in the LR image while the HR image is being recovered. Therefore, the SR techniques are proven to be restrictive for usage in the practical applications~\cite{baker2002limits}. SR could be used for many different applications such as satellite and aerial imaging~\cite{thornton2006sub}, medical image processing~\cite{8471089}, infrared imaging~\cite{zhao2015novel}, improvement of text, sign and license plate~\cite{banerjee2008super}, and finger prints~\cite{cui2004iris}. 
	
\begin{figure}
		\begin{center}
			\includegraphics[width=7.5cm]{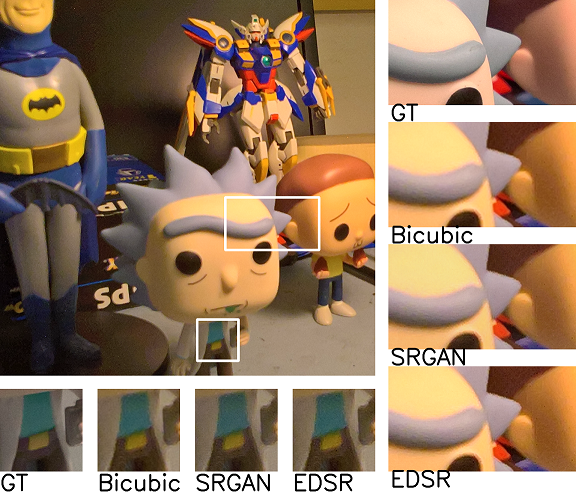}
		\end{center}
		\vspace{-.2cm}
		\caption{Super Resolution Process on an image from ImagePairs dataset using bicubic, SRGAN~\cite{ledig2017photo} and EDSR~\cite{Lim_2017_CVPR_Workshops} methods.}
		\label{fig:example}
		\vspace{-.25cm}
	\end{figure}

	Figure~\ref{fig:example} shows an example of single-image SR process where the recovered HR image is 4 times larger that its LR input image. We show the result of different super resolution techniques in this figure. If the technique fails to recover adequate detail from the LR input, the output will be blurry without sharp edges.
	
	The SR problem has been studied comprehensively in the past~\cite{tian2011survey, nasrollahi2014super} and many machine learning techniques has been proposed to solve this problem. Examples would include Bayesian~\cite{tipping2002bayesian}, steering kernel regression~\cite{zhang2012single}, adaptive Wiener filter~\cite{hardie2007fast}, neighbor embedding~\cite{gao2012image, bevilacqua2012low}, matching~\cite{SR_sun_hays_2012} and example-based~\cite{freeman2002example} methods.
	
	Deep learning techniques have been proven a success in many areas of computer vision. This involves application of deep learning techniques by the lead image restoration researcher to solve SR~\cite{Huang-CVPR-2015,LedigTHCATTWS16, Kim_2016_VDSR, wang2015deep, liu2016robust, liu2016learning, NIPS2014_5423, dong2014learning}. Because of the nature of deep the learning networks, being a multi-layered feature extraction cascade~\cite{deng2014deep}, more data is required in order to train these complex methods~\cite{schmidhuber2015deep}. 
	
	As proven, the input data itself plays a significant role in the machine learning processes~\cite{banko2001scaling, zhu2012we}, especially deep learning approaches which are data hungry. Hence, the process of gathering data and its formation may be equally as vital to solving the machine learning problem as the technique used. 
	The sole purpose of SR is not to upscale or to increase the number of pixels in an image, but to increase the quality of it as closely to an image with the target resolution as possible.
	An example would be capturing a photo using a cellphone with a $5MP$ front facing camera and a $20MP$ rear facing camera where a 2X SR technique applied to the front facing camera will make it $20MP$. This is an attempt to increase the number of pixel from $5MP$ to $20MP$ while expecting an increase in the the quality the output image similar to that of the high quality rear facing camera. An example is presented in Fig.~\ref{fig:example}, where the same scene was photographed with a $5MP$ camera and $20.1MP$ camera in the same lighting condition. The same part of the image was cropped to show the nature of the difference in the quality of the images (ground truth vs. bicubic). This shows that maintaining the $20MP$ quality of the SR technique output requires SR, noise cancellation, image sharpening and even color correction to some extend while the-state-of-the-art methods such as SRGAN~\cite{ledig2017photo} and EDSR~\cite{Lim_2017_CVPR_Workshops} fail to do so as seen in Fig.~\ref{fig:example}. We believe that the main reason for failure of these methods is lack of realistic training data that we focus on this paper.   
	
	A more complex version of this task could be Image Signal Processing (ISP) pipeline with various stages including denoising~\cite{buades2005non,zhang2017beyond}, demosaicing~\cite{li2008image}, gamma correction, white balancing~\cite{joze2013exemplar,joze2012role} and so on. ISP pipeline has to be tuned by camera experts for a relatively long time before it can be used in the commercial cameras. Domain knowledge such as optics, mechanics of the cameras, electronics and human perception of colors and contrast are necessary in this tuning process. Replacing this highly skilled and tedious tuning process with a deep neural network is a recent research direction in computational photography \cite{Schwartz2019DeepISP,Ratnasingam2019,Guidance_Network}.
%	\begin{figure}
%		\begin{center}
%			\includegraphics[width=4.1cm]{low_res_cut}
%			\includegraphics[width=4.1cm]{hi_res_cut}
%		\end{center}
%		\caption{A scene was captured with a 5MP camera and 20MP camera in the same lighting condition. The same part of the images was cropped and presented for comparison. (The LR image has been resized for the purpose of visualization)}
%		\label{fig:low-high-res}
%	\end{figure}
	Current datasets~\cite{yang2014single, Agustsson_2017_CVPR_Workshops} widely used for training SR models increase the number of pixel without taking the quality of the image into consideration. 
	The new data acquisition technique proposed herein may be used for SR, noise cancellation and quality enhancement techniques. A dataset of 11,421 pairs of LR-HR images is presented which was used to solve the SR problem. We use a beam-splitter to capture the same scene by two cameras: LR and HR. The proposed device can capture the same scene by two cameras, there still have a different perspective due to the different focal lenses, but we solve it by local alignment technique. Since we also release the raw images, this large-scale dataset could be used for other tasks such as ISP generation. To our knowledge, this is the most complete dataset for SR, ISP and image quality enhancement with far more real LR-HR images compared to existing dataset for SR and ISP task. This dataset is more than $10\times$ larger than current SR dataset while it includes real LR-HR pairs and more than $2\times$ larger than current ISP dataset while it includes diverse scenes. The benchmark result shows how the new dataset can be successfully used to significantly improve the quality of real-world image super resolution.
%	
	%The outline of the paper is as follows: related work discussed in sections~\ref{sec:two}. The proposed data acquisition technique discussed in section~\ref{sec:three}. In section~\ref{sec:four}, ImagePairs data set and its generation process is introduced. The SR neural network model trained using proposed data set is presented and compared to existing SR models in section~\ref{sec:six}. Finally, the paper is concluded and the future work is discussed in section~\ref{sec:seven}.
%	
\begin{table*}[h]
\begin{center}
\begin{tabular}{ l c c c c c}
%\hlineB{3}
%\cline{3-7}
%\cline{2-3} 
\toprule
Data Set & Size & Main purpose & HR Resolution & LR generation & Raw  \\
\hline
Set5~\cite{bevilacqua2012low}, Set14~\cite{zhao2015loss}, Urban100~\cite{huang2015single} & 5/14/100 & SR & $512\times512$  & down-sample  HR & No \\
 The Berkeley segmentation~\cite{martin2001database} & 200 & Segmentation & $481\times321$  & down-sample  HR & No \\
 DIV2K~\cite{Agustsson_2017_CVPR_Workshops} & 1000 & SR & $2048 \times 1080$ & down-sample  HR & No \\
 %Flicker2K~\cite{Lim_2017_CVPR_Workshops} & & & \\ 
 See-In-the-Dark (SID) \cite{ChenSid_CVPR2018} & 5094 & Low-Light & $4240\times2832$ & - & Yes \\
 Samsung S7 \cite{Schwartz2019DeepISP} & 110 & ISP &  $4032\times3024$  & - & Yes \\
 
 RealSR ~\cite{cai2019toward} & 595 & SR & $3500\times700$ & Real & Yes \\ 
 
 ImagePairs (Proposed) & 11421 & SR & $3504\times2332$  & Real & Yes \\
\bottomrule
\end{tabular}
\vspace{.1cm}
\caption{Compression between proposed dataset to current datasets used for its task.}
\end{center}
\label{tbl:dataset}
\vspace{-.25cm}
\end{table*}
	\section{Related Works} \label{sec:two}  
	In recent years, the core of image SR methods has shifted towards machine learning, mainly the machine learning techniques and the datasets. 
	Herein, a brief description is given on the single image SR methods and learning-based ISP methods as well as their common datasets. There are also multiple image SR methods~\cite{tipping2002bayesian,borman1998super,farsiu2004fast, dong2014learning} which are not the main focus of this paper. More comprehensive SR methods descriptions may be found in \cite{tian2011survey, nasrollahi2014super}. 
	
	\subsection{SR methods}
	\textbf{Interpolation based:} The early SR methods are known as interpolation based methods where new pixels are estimated by interpolating given pixels. This is the easiest way to updates the image resolution. Examples include Nearest Neighbor interpolation, Bilinear interpolation and Bicubic interpolation which uses 1, 4 and 16 neighbor pixels respectively to compute the value of new pixels. These methods are wildly in use in image resizing.
	
	\textbf{Patch based:} More recent SR methods rely on machine learning techniques to learn the relation between patches of HR image and patches of LR images. These methods are referred to as patch-based methods in some literature~\cite{yang2014single, wang2005patch} and Exemplar-based in other~\cite{freeman2002example, glasner2009super}. Unlike the first class of methods, these methods need training data in order to train their models. These training data are usually pairs or corresponding LR and HR images. The training dataset is further discussed in subsection~\ref{subsec:datasets}.
	Depending on the source of a training patch, the corresponding
	method for patch based SR may be categorized into two main categories: external or internal. 
	
	\textbf{External methods} the external method uses a variety of learning algorithms to learn the LR-HR mapping
	from a large database of LR-HR image pairs. These include
	nearest neighbor~\cite{freeman2002example}, kernel ridge regression~\cite{kim2010single}, sparse
	coding~\cite{yang2012coupled} and convolutional
	neural networks~\cite{dong2014learning}.
	
	\textbf{Internal Methods} the internal method on the other hand assumes that patches of a natural image recurs within and across scales of the same image~\cite{barnsley2014fractals}. Therefore, it makes an attempt to search for a HR patch within a LR image with different scales.
	Glasner et al.~\cite{glasner2009super} united the classical and example-based SR by exploiting the patch recurrence within and across image scales. Freedman and Fattal~\cite{freedman2011image} gained computational speed-up by showing that self-similar patches can often be found in limited spatial neighborhoods. Yang et al.~\cite{yang2014single} reﬁned this notion further to seek self-similar patches in extremely localized neighborhoods, and performed ﬁrst-order regression. Michaeli and Irani~\cite{michaeli2013nonparametric} used self-similarity to jointly recover the blur kernel and the HR image. Singh et al.~\cite{huang2015single} used the self-similarity principle for super-resolving noisy images.  
	
	With the success of convolution neural networks, many internal patch-based SR methods were proposed which outperform the prior methods. As an example, SRGAN~\cite{ledig2017photo} used a generative adversarial network (GAN)~\cite{goodfellow2014generative} for this task that trained by perceptual loss function consisting of an adversarial loss and a content loss.
	The residual dense networks (RDN)~\cite{zhang2018residual} exploited the hierarchical features from all the convolutional layers. EDSR~\cite{Lim_2017_CVPR_Workshops} did a performance improvement by removing unnecessary modules in conventional residual networks. WDSR~\cite{yu2018wide} introduced a linear low-rank convolution in order to further widen activation without computational overhead.

	\subsection{ISP Methods}
	Image Signal Processing (ISP) pipeline is a method used to convert an raw image into a digital form in order to get an enhanced image. This consists of various stages including denoising~\cite{buades2005non,zhang2017beyond}, demosaicing~\cite{li2008image}, gamma correction, white balancing~\cite{joze2012exemplar,drew2014zeta} and so on. Currently, this pipeline has to be tuned by camera experts for a long period of time for each new camera.
	Replacing the expert-tuned ISP with a fully automatic method has been done with few recent methods approach by training an end-to-end deep neural network~\cite{Schwartz2019DeepISP,Ratnasingam2019,Guidance_Network}. Schwartz et al.~\cite{Schwartz2019DeepISP} released a data set, named Samsung S7 data set, contains RAW and RGB image pairs with both short and medium exposures. They design a network that first processes the image locally then globally. Ratnasingam~\cite{Ratnasingam2019} replicates the steps of a full ISP with a group of sub networks and achieves the-state-of-the-art result by training and testing on a set of synthetic images. Liang et al.~\cite{Guidance_Network} used 4 sequential u-nets in order to solve this problem. They claimed that the same network can be used for en-lighting extreme low light images. 
	
	\subsection{SR and ISP Datasets} \label{subsec:datasets}
	
	SR dataset includes pairs of HR and LR images. Most existing datasets generate an LR image from the corresponding HR image by sub-sampling the image using various settings. Here the HR images are also called ground truth as the final goal of SR methods is to retrieve them from LR images. Therefore, SR dataset includes sets of HR images or ground truths and settings to generate LR image from HR images. Here is a list of common SR datasets:
	
	\begin{enumerate}
	\vspace{-.2cm}
	    \setlength\itemsep{0em}
		\item 
		The Berkeley segmentation dataset~\cite{martin2001database} is one of the first datasets used for single image SR~\cite{sun2010context, freedman2011image,glasner2009super}. It includes 200 professional photographic style images of $481\times321$ pixels with a diverse content.
		\item 
		Yang et. al.~\cite{yang2014single} proposed a benchmark for single image SR which includes The Berkeley segmentation dataset as well as a second set containing 29 undistorted high-quality images from the LIVE1 dataset~\cite{sheikh2006statistical} , ranging from $720\times480$ to $768\times512$ pixels. 	
		Huang et. al.~\cite{Huang-CVPR-2015} added 100 urban high resolution images from flicker100 with a variety of real-world
		structures to this benchmark, in order to focus more on man made object.
		\item DIV2K dataset~\cite{Agustsson_2017_CVPR_Workshops} has introduced a new challenge for single image SR. This database include 1000 images of diverse contents with train/test/validation split as 90/10/10.
		\item RealSR dataset~\cite{cai2019toward} captured images of the same scene using ﬁxed DSLR cameras with different focal lengths. The focal length changes can capture ﬁner details of the scene. This way, HR and LR image pairs on different scales can be collected with a registration algorithm. This dataset includes 595 LR/HR pairs of indoor and outdoor scenes.   
		\vspace{-.2cm}
	\end{enumerate}
	
	There are also few standard benchmark datasets, Set5~\cite{bevilacqua2012low}, Set14~\cite{zhao2015loss}, and Urban100~\cite{huang2015single} commonly used for performance comparison. These datasets include 5 ,14 and 100 images, respectively.
	Apart from RealSR~\cite{cai2019toward}, all other datasets do not include LR images so the LR image should generate synthetically from corresponding HR image. There are several ways to generate LR test images from HR images  (the ground truth)~\cite{shan2008fast, timofte2013anchored,sun2008image} such that the generated LR test images may be numerically different. One common way to achieve this is to generate a LR image in a Gaussian blur kernel to down-sample the HR image using a noise term~\cite{irani1991improving,kim2010single,yang2014single}. The parameter for this task will be $s$ as scale factor, $\alpha$ for Gaussian kernel and $\epsilon$ for noise factor. There are other datasets dedicated to image enhancements such as MIT5K~\cite{fivek} and DPED~\cite{ignatov2017dslr}.  
	MIT5K~\cite{fivek} includes 5,000 photographs taken with SLR cameras, each image retouched by professionals to achieve visually pleasing renditions.
	DPED~\cite{ignatov2017dslr, ignatov2019ntire} consists of photos taken synchronously in the wild by three smartphones and one DSLR camera. The smartphone images were aligned with DSLR images  to extract $100\times100$ patches for CNN training including 139K, 160K and 162K pairs for each settings. This dataset was used in a challenge on image enhancement~\cite{ignatov2019ntire} as well as a challenge on RAW to RGB Mapping~\cite{ignatov2019aim}. 
	
	There are not many publicly available ISP dataset which requires raw image as well as generated image from that. Here we describe two datasets that were used for ISP.  

	\begin{enumerate}
		\vspace{-.1cm}
	 \setlength\itemsep{0em}
\item \textbf{See-In-the-Dark (SID)}: proposed by Chen et al. \cite{ChenSid_CVPR2018}, is a Raw-RGB dataset captured in extreme low-light where each short-exposure raw image is paired with its long-exposure RGB counterpart for training and testing \cite{Zamir2019}. Images in this dataset were captured using two cameras: Sony $\alpha7S II$ and Fujifilm X-T2, each subset contains about 2500 images, with about $20\%$ as test set. The raw format of Sony subset is the traditional 4-channel Bayer pattern that of Fuji subset is XTrans format with 9 channels. Beside raw and RGB data, their exposure times are provided alongside. 
    
    \item \textbf{Samsung S7}: captured by Schwartz et al. \cite{Schwartz2019DeepISP}, contains 110 different RAW-RGB pairs, with train/test/validation split of 90/10/10. Different to the SID dataset, this one does not provide related camera properties such as the exposure time associated with the image pairs. The raw format here is also the traditional 4-channel Bayer pattern.	\vspace{-.1cm}
\end{enumerate}

	Current SR methods as well as learning based ISP methods are mainly focused on their learning process as mentioned before. Different machine learning techniques have been applied to these problems and recent efforts have involved training different deep neural network models. Comparing to datasets for popular computer vision tasks such as image classification~\cite{deng2009imagenet},  detection~\cite{Everingham15, lin2014microsoft}, segmentation~\cite{lin2014microsoft}, video classification~\cite{kay2017kinetics} and sign language recognition~\cite{vaezijoze2019ms-asl}, there is an obvious lack of large realistic dataset for SR and ISP tasks despite of the potential to produce significant result by neural network techniques.
	Table~\ref{tbl:dataset} shows all these datasets currently used for SR and ISP tasks and their specification compared to our proposed dataset ImagePairs. Our proposed dataset is not only at least 10 times larger than other SR datasets and 2 times from other ISP datasets, but also has real LR-HR images and includes raw images which could be used for other tasks. 
	
	%new deep learning work:
	%~\cite{Huang-CVPR-2015}
	%~\cite{LedigTHCATTWS16}
	%~\cite{Kim_2016_VDSR}
	%~\cite{wang2015deep}
	%~\cite{liu2016robust}
	%~\cite{liu2016learning}
	%~\cite{NIPS2014_5423}
	
	%single image not deep learning:
	%~\cite{zhao2003wavelet}
	%~\cite{zhang2012single}
	%~\cite{gao2012image}
	%~\cite{hardie2007fast}
	%~\cite{zhang2012single}
	%~\cite{bevilacqua2012low}
	%~\cite{SR_sun_hays_2012}

\section{Data Acquisition} \label{sec:three}

\subsection{Hardware Design}
 The high resolution camera used had a $20.1MP$, 1/2.4” format CMOS image sensor supporting $5344(H) \times 3752(V)$ frame capture, $1.12\mu m$ pixel size, and lens focal length of $f=4.418mm$ (F/1.94), providing a $68.2^{\circ} \times 50.9^{\circ}$ field of view (FOV). The camera also featured bidirectional auto-focus (open loop VCM) and 2-axis optical image stabilization (closed loop VCM) capability.

The lower resolution fixed-focus camera used had a similar FOV with approximately half the angular pixel resolution. it also featured a $5MP$, 1/4” format CMOS image sensor supporting $2588 \times 1944$ frame capture, $1.4\mu m$ pixel size, and lens focal length $f=2.9mm$ (F/2.4), providing a $64^{\circ}(H) x 50.3^{\circ}(V)$ FOV. Table~\ref{table:camera} shows the specifications for these cameras.

	\begin{table}[t]
		\begin{center}
			\begin{tabular}{l|c|c}
			   \toprule
				Camera & Low-resolution & High-resolution \\
				\hline
				Image sensor format & 1/4”  & 1/2.4” \\
				pixel size & $1.4\mu m$ & $1.12\mu m$ \\
				Resolution & 5MP & 20.1MP \\
				FOV (H,V) & $64^{\circ}$, $50.3^{\circ}$ &  $68.2^{\circ}$, $50.9^{\circ}$\\
				Lens focal length & $2.9mm$ & $4.418mm$ \\
				Focus & fixed-focus & auto-focus \\
				\bottomrule
			\end{tabular}
		\end{center}
		\vspace{-.2cm}
		\caption{Camera Specifications.}
		\label{table:camera}
		\vspace{-.2cm}
	\end{table}

	\begin{figure}[b]
	    \vspace{-.25cm}
		\begin{center}
			\includegraphics[width=5cm]{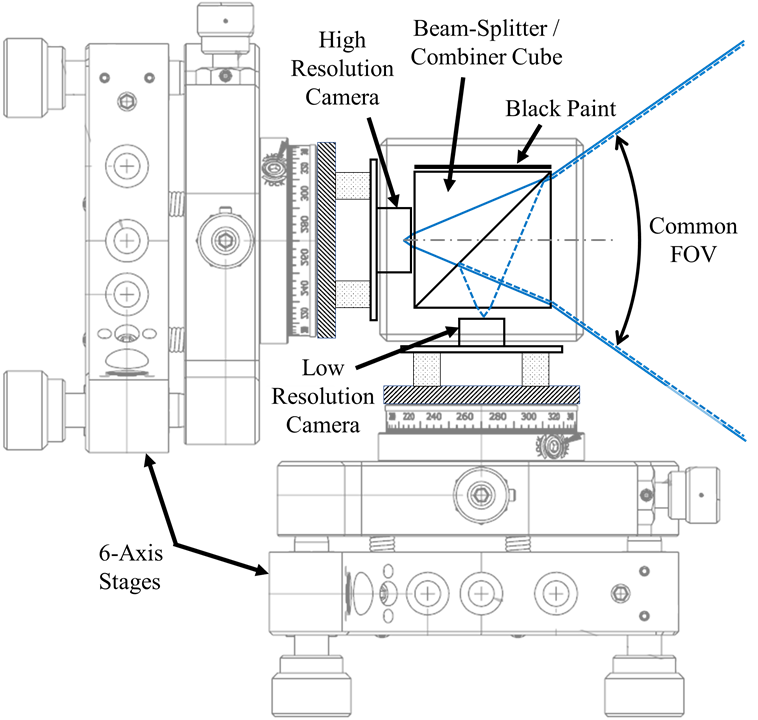}
			%\fbox{\rule{0pt}{2in} \rule{.9\linewidth}{0pt}}
		\caption{Opto-mechanical layout of dual camera combiner, showing high resolution camera (transmission path) and low resolution camera (reflective path) optically aligned at nodal points and with overlapping FOV pointing angle.}
		\label{fig:dualcamsetup}
		\end{center}
	\end{figure}
	
		\begin{figure}[b]
		\vspace{-.35cm}
		\begin{center}
			\includegraphics[width=3.65cm]{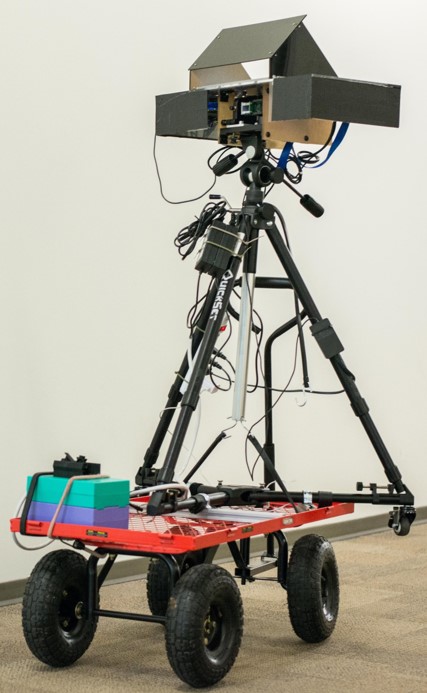}
			%\fbox{\rule{0pt}{2in} \rule{.9\linewidth}{0pt}}
		\end{center}
		\vspace{-.25cm}
		\caption{ The data acquisition device install on a tripod while the trolley is used for outdoor manoeuvre.}
		\label{fig:camera}
	\end{figure}
	
	\begin{figure}[t]
		\vspace{-.1cm}
		\begin{center}
			\includegraphics[width=5.5cm]{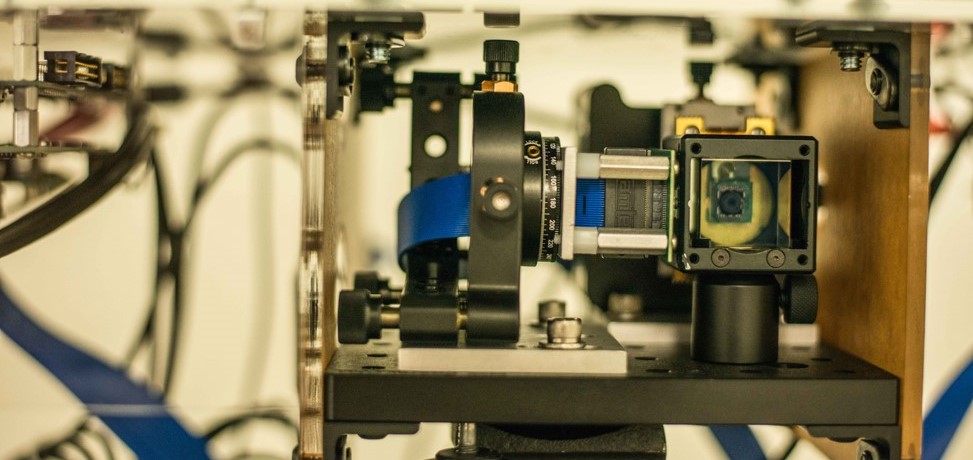}
		\caption{Two camera setup}
		\label{fig:camera_setup}
		\end{center}
		\vspace{-.5cm}
	\end{figure}

In order to simultaneously capture frames on both cameras with a common perspective, the FOVs of both cameras are combined using a Thorlabs BS013 50/50 non-polarizing beam-splitter cube. They are then aligned such that pointing angle of the optical axes are at far distance and entrance pupils at each camera (nodes)are  at near distance.
%appear to overlay while being mounted at orthogonal transmission and folded optical paths. 
The high resolution camera, placed behind the combiner cube in the transmission optical path, is mounted on a Thorlabs K6XS 6-axis stage so that the $x$ and $y$ position of the entrance pupil is centered with the cube and the $z$ position in close proximity. The tip and tilt of camera image center field pointing angle is aligned with a target at distance while rotation about camera optical axis is aligned by matching pixel row(s) with a horizontal line target. Fig.~\ref{fig:dualcamsetup} illustrates the opto-mechanical layout of the dual camera combiner. The low resolution camera is placed behind the combiner cube in the lateral $90^{\circ}$ folded optical path and also mounted on a 6-axis stage. It is then aligned in $x$, $y$ and $z$ such that entrance pupil optically overlaps that of the high resolution camera. The tip/tilt pointing angle as well as camera rotation about optical axis may be adjusted so as to achieve similar scene capture. In order to refine the overlap toward pixel accuracy, a live capture tool displays the absolute difference of camera frame image content between cameras such that center pointing and rotation leveling may be adjusted with high sensitivity. Any spatial and angular offsets may be substantially nulled by mechanically locking the camera in position. The unused combiner optical path is painted with carbon black to limit image contrast loss due to scatter. The opto-mechanical layout of dual camera combiner is illustrated at figure ~\ref{fig:camera_setup}. 

The proposed device can capture the same scene by two different cameras. The two cameras have a difference in perspective due to the different focal lenses which was solved by a local alignment technique described in section~\ref{sec:four}. Furthermore, the two camera sensors get half of the light because of 50/50 split with poorer image quality mainly on low-resolution camera.

\subsection{Software Design}
	A data capturing software was developed to connect to both cameras, allowing them to synchronize with each other. The software may capture photo from both cameras at the same time as well as adjusting camera parameters such as gain, exposure and lens position for the HR camera. The raw data was stored for each camera, allowing later use of the arbitrary ISP. For each camera, all the meta data was stored on a file including the image category selected by the photographer.
	Figures~\ref{fig:camera} shows the data acquisition device installed on a tripod while the trolley is used for outdoor maneuvering.
	
\section{ImagePairs Dataset} \label{sec:four}
	
	The dataset was called ImagePairs as it includs pairs of images of the exact scene using two different cameras. Images are either LR or HR where the HR image is twice as big in each dimensions as the corresponding LR image; all LR images are $1752\times1166$ pixels and HR images are $3504\times2332$ pixels. Unlike other real world datasets, we do not use zooming levels or scaling factor to increase the number of pairs so each pair corresponds to a separate scene. This means that we captures 11,421 distinct scenes with the device which generates 11,421 image pairs.  
	
\begin{figure}[h]
	\begin{center}
	   \setlength{\tabcolsep}{2pt} % Default value: 6pt
		\begin{tabular}{cc}
			\includegraphics[width=4.1cm]{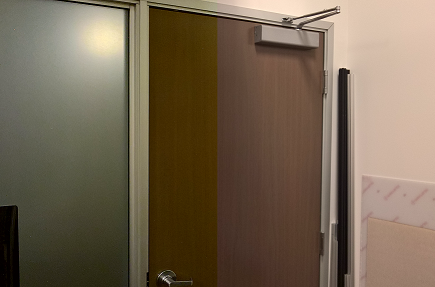} &
			\includegraphics[width=4.1cm]{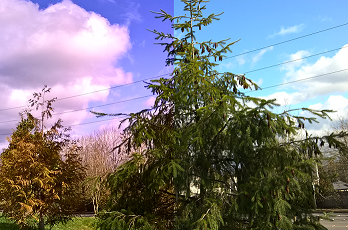} \\
			\includegraphics[width=4.1cm]{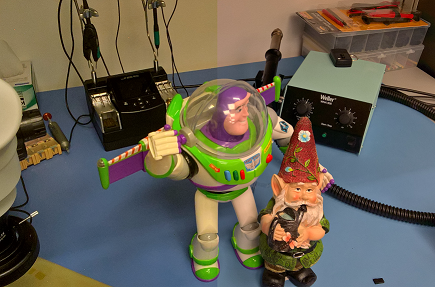} &
			\includegraphics[width=4.1cm]{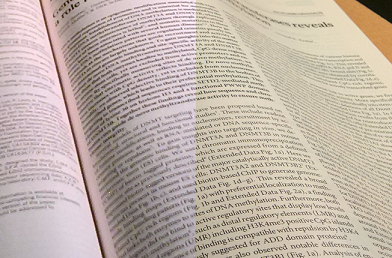} \\
			\includegraphics[width=4.1cm]{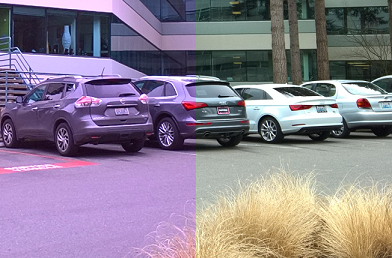} &
			\includegraphics[width=4.1cm]{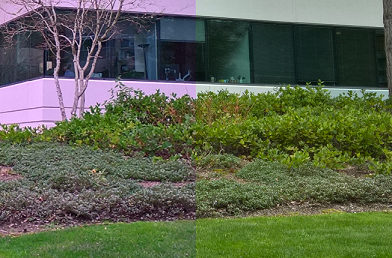} \\
		\end{tabular}
		\caption{Sample images from ImagePairs dataset. Each image divided by half horizontally to show LR on the left and HR on the right.}
		\label{fig:ds}
		\vspace{-.45cm}
	\end{center}
\end{figure}

	For each image pair, the meta data such as gain, exposure, lens position and scene categories were stored. Each image pair was assigned to a category which may later be used for training purposes. These categories include Document, Board, Office, Face, Car, Tree, Sky, Object, Night and Outdoor. The pairs are later divided in two sets of train and test, each including 8591 and 2830 image pairs, respectively. %~\footnote{ \url{www.microsoft.com/en-us/research/project/imagepairs} }. 
	The two cameras have a difference in perspective due to the different focal lenses. Therefore, in order to generate pairs corresponding to each other in pixel level, the following steps were applied: (1) ISP (2) image undistortion (3) pair alignment (4) margin cropping.  Figure~\ref{fig:ds} illustrates diverse samples from proposed dataset after the final alignments. In order to show the accuracy of pixel-by-pixel alignment,  each sample image is divided by half horizontally to show LR at left and HR at right in Fig.~\ref{fig:ds}.
	
	\textbf{ISP : }
	The original images were stored in the raw format. The first step was to convert the raw data to color images, using a full-stack powerful ISP for both LR and HR. Since we have access to the raw data, the ISP can be replaced with a different one or a simple linear one to ignore the non-linearity in the pipeline.
	
\begin{figure}[t]
		\begin{center}
			\scalebox{-1}[1]{\includegraphics[width=3.9cm]{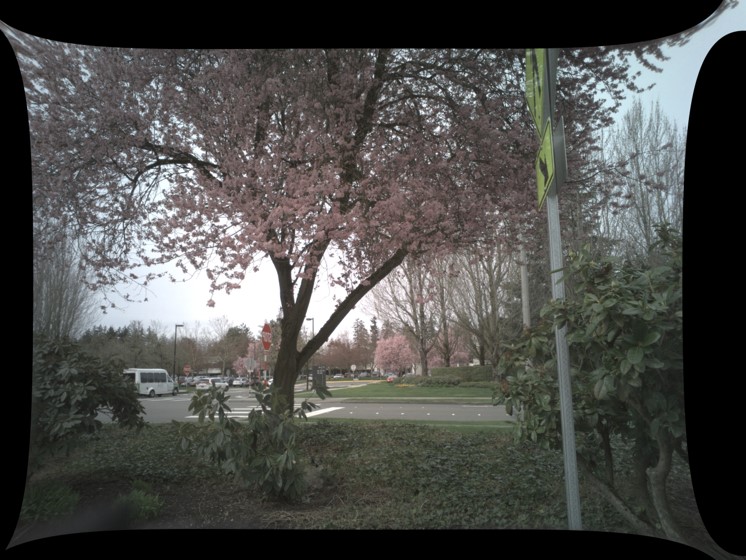}}
			\includegraphics[width=4.2cm]{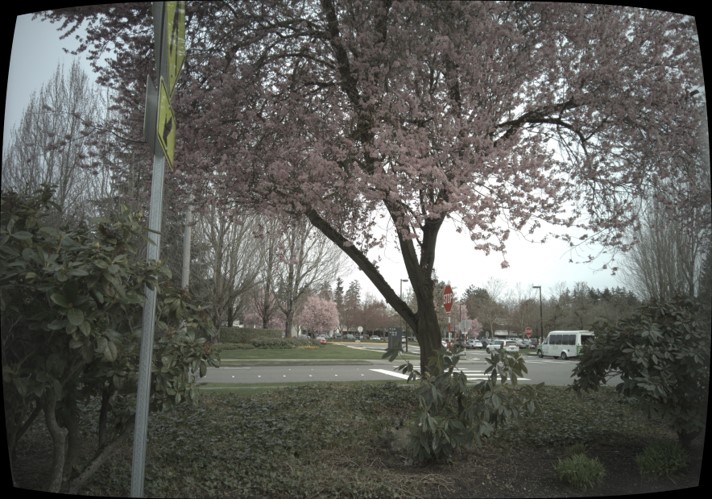}
		\end{center}
		\vspace{-.25cm}
		\caption{Undistorted image of HR at right and LR at left.} % after calibration.}
		\label{fig:distotion}
		 \vspace{-.35cm}
	\end{figure}

	\textbf{Image Undistortion :}
	CMOS cameras introduce a lot of distortion to images. Two major distortions are radial distortion and tangential distortion. In radial distortion, straight lines will appear curved while in the tangential distortion the lens is not aligned perfectly parallel to the imaging plane.
	To overcome these distortions in both LR and HR images, we calibrated both cameras by capturing several checkerboard images. These images were later used to solve a simple model for radial and tangential distortions~\cite{opencv_undistort}. Figure~\ref{fig:distotion} shows the un-distorted image for both LR and HR images.
	
	\textbf{Alignment :}
	We use two steps in order to align the LR and HR images. First we try to globally match two images using image registration technique specifically homography transformation. Although now HR and LR image are globally aligned but they may not be aligned pixel by pixel due to some geometry constrains. So as the second step, we use a 10 by 10 grid for LR image and do a local search to find the best match on HR image for that grid. Lastly, we use matching position for grids on HR image to warp the LR image so that the LR and HR are globally and locally matched to each other.      
	
\begin{figure}[b]
 \vspace{-.50cm}
\begin{center}
    \includegraphics[width=8.5cm]{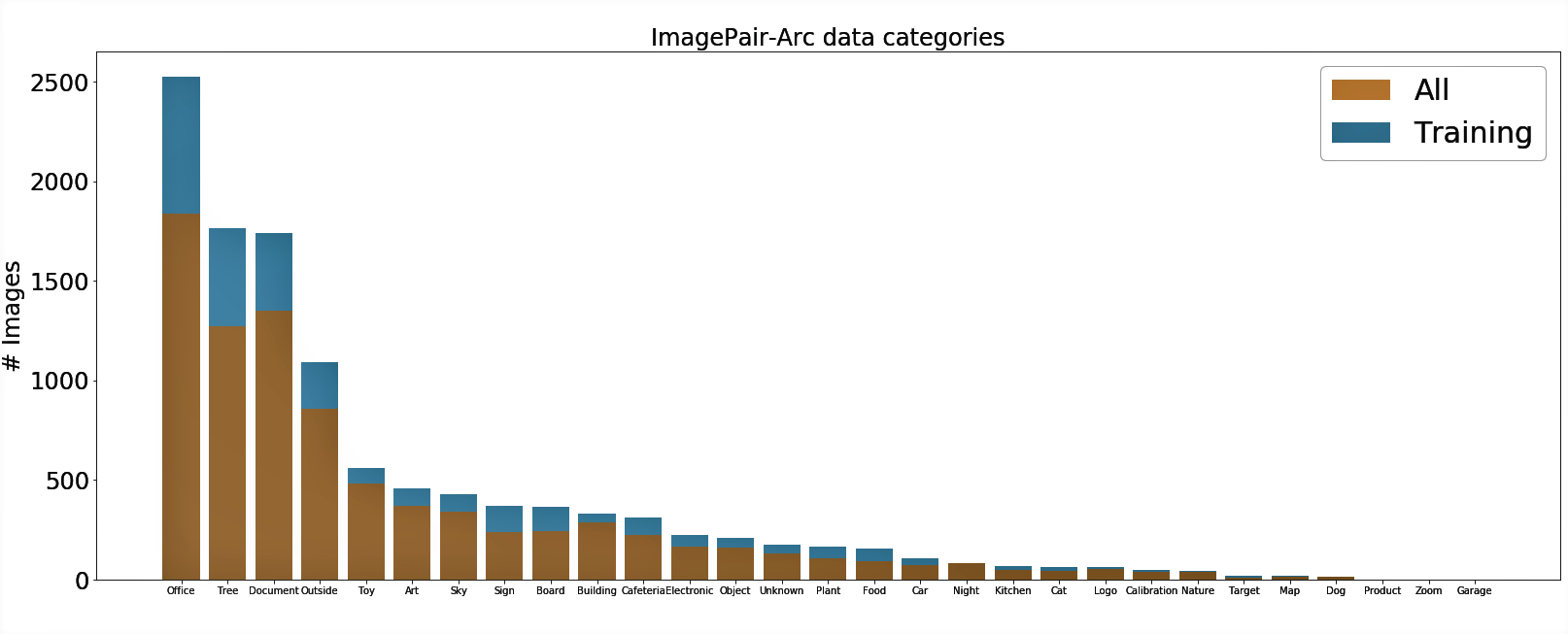}
\vspace{-.40cm}
    \caption{Frequency of ImagePairs train/test categories.  }
    \label{fig:stats}
\end{center}
\vspace{-.25cm}
\end{figure}

	\textbf{Margin Crop :}
    Although the images were aligned globally and locally, the borders are not as aligned as we expected, possibly due to differences in the camera specifications. Therefore, $10\%$ of border from each image was removed, 
	resulting in a change in the resolution of both LR and HR images; $1752\times1166$ pixels and $3504\times2332$ pixels respectively. 
For each image (LR or HR) we also stored meta data which is 
analogue gain, digital gain, exposure time, lens position and scene category. The scene category which is selected by the photographer includes Office, Document, Tree, Outside, Toy, Sky, Sign, Art, Building, Night, etc. Figure~\ref{fig:stats} illustrates the frequency of each categories for ImagePairs train/test sets. 

At this point, the ImagePairs consists of a large dataset of HR-LR images, allowing the easy application of patch-base algorithms. Random patches can pick from LR and the corresponding HR patches. Since the correspondence is pixel by pixel, there is no need to search for similar patches in different scales. Additionally, the ground truth (HR) has 4 times more pixels, is sharper and less noisy compared to the LR images, hence an increased image quality. 

\begin{figure*}
    \vspace{-.25cm}
	\begin{center}
		\begin{tabular}{C{2.2cm}C{2.2cm}C{2.2cm}C{2.2cm}C{2.2cm}c}
			Bicubic & RDN~\cite{zhang2018residual} & SRGAN~\cite{ledig2017photo} & EDSR~\cite{Lim_2017_CVPR_Workshops} & WDSR~\cite{yu2018wide} & Proposed \\
			\multicolumn{6}{c}{\includegraphics[width=15.75cm]{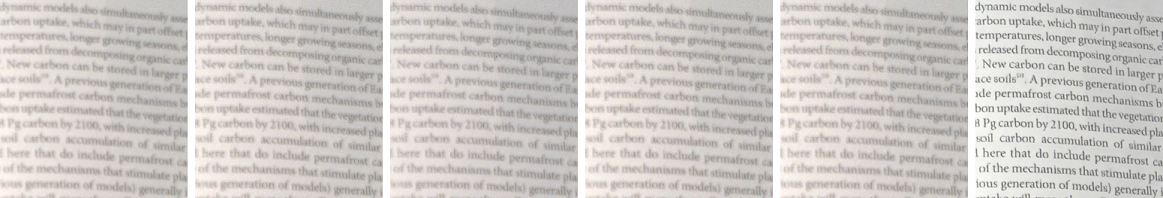}} \\
			\multicolumn{6}{c}{\includegraphics[width=15.75cm]{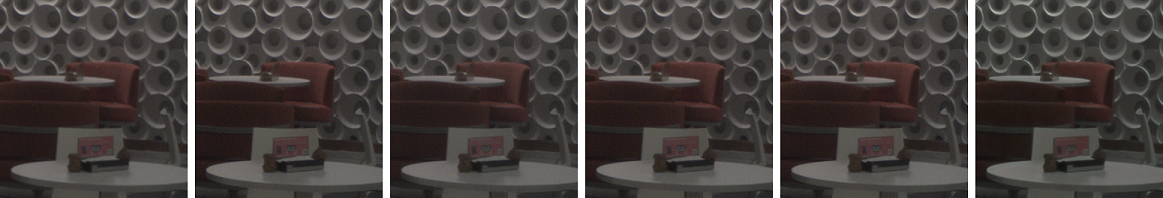}} \\
			\multicolumn{6}{c}{\includegraphics[width=15.75cm]{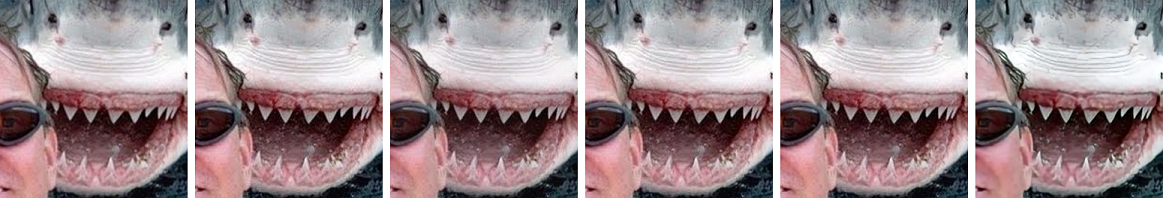}} \\
			\multicolumn{6}{c}{\includegraphics[width=15.75cm]{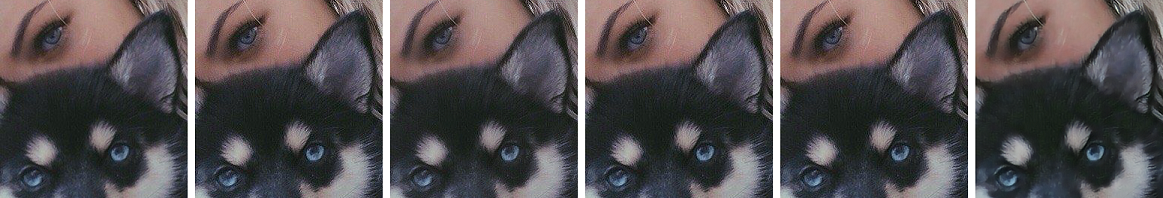}} \\
		\end{tabular}
		\caption{Qualitative compassion of the-state-of-the-art super resolution methods train on DIV2K~\cite{Agustsson_2017_CVPR_Workshops} dataset set compare to simple network trained on ImagePairs dataset. The first two images are from ImagePairs test set and the next two are external images.}
		\label{fig:compare}
	\end{center}
	\vspace{-.4cm}
\end{figure*}

\section{Experimental Results} \label{sec:six}

\subsection{Realistic Super Resolution}
Before running a benchmark for state-of-the-art SR methods, we need to see their performance when trained on current SR datasets. As mentioned before, a real LR image usually has many other artifacts as it is captured with a weaker camera. We train a basic generative adveral network (GAN) model which includes 10 convolution layers for generator and a U-Net with 10 convolution/deconvolution for discriminator network with the proposed dataset. The sole reason of this experiment is to see if current state-of-the-art methods trained on synthetic images can outperform our simple method training on real images or not. Figure~\ref{fig:compare} shows the performance of this method compared to the super resolution methods: SRGAN~\cite{ledig2017photo}, EDSR~\cite{Lim_2017_CVPR_Workshops}, WDSR~\cite{yu2018wide} and RDN~\cite{zhang2018residual} trained on DIV2K dataset~\cite{Agustsson_2017_CVPR_Workshops}. The first two images are from ImagePairs test set and the next two images are from real-world LR images from the internet. As expected, these methods only increase the pixel and do not effect image artifacts like noise and color temperature. Our method trained on ImagePairs dataset does well for test images from the dataset and real-world LR images.

\subsection{Super Resolution Benchmark}
\begin{figure*}
    \vspace{-.25cm}
	\begin{center}
		\begin{tabular}{C{3.0cm}C{3.0cm}C{3.0cm}C{3.0cm}c}
			HR & Bicubic & WDSR~\cite{yu2018wide}  & EDSR~\cite{Lim_2017_CVPR_Workshops} &  SRGAN~\cite{ledig2017photo} \\ 
			\multicolumn{5}{c}{\includegraphics[width=17cm]{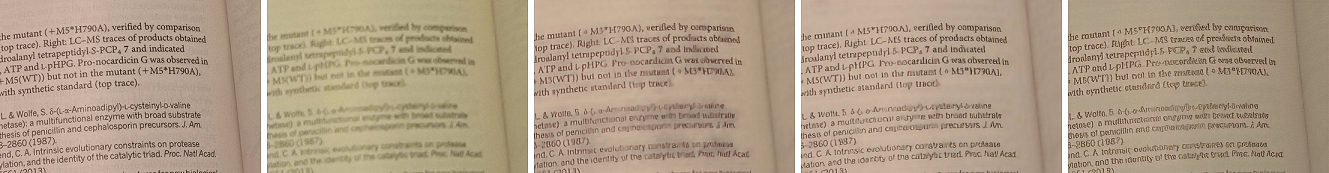}} \\
			\multicolumn{5}{c}{\includegraphics[width=17cm]{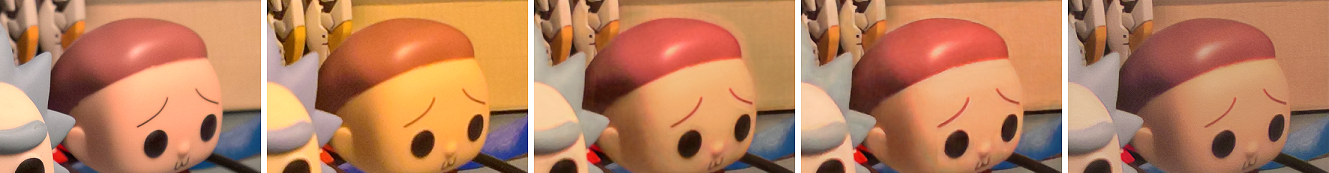}}
		\end{tabular}
		\caption{Qualitative comparison of the-state-of-the-art super resolution methods train on proposed dataset.}
		\label{fig:compare_final}
	\end{center}
\end{figure*}

We trained three 2X super resolution methods on ImagePairs train set including SRGAN~\cite{ledig2017photo}, EDSR~\cite{Lim_2017_CVPR_Workshops} and WDSR~\cite{yu2018wide} by using their model implementation by~\cite{krasser2018git,cardinale2018isr}. All SR methods trained using LR-HR rgb images and we do not use raw images as input. We use same patch size of $128\times128$ for HR images and batch size equal to $16$ for all training. All methods are trained for $150K$ iterations. For evaluation, we run trained models on centered quarter of cropped images of Imagepairs test set.  Table~\ref{tab:sr_comapre} reports the peak signal-to-noise ratio (PSNR) and the structural similarity (SSIM)~\cite{wang2004image} for trained model on ImagePairs as well as model trained on DIV2K dataset with similar parameters. As we discussed before, the PSNR and SSIM for methods trained on DIV2K is comparable with bicubic method. In some cases, they perform worst than bicubic since noise could boost with some SR methods. On the other hand, when we trained the same models with proposed ImagePairs dataset, all methods outperform their PNSR. SRGAN~\cite{ledig2017photo} and EDSR~\cite{Lim_2017_CVPR_Workshops} is doing a good job in noise cancellation and outperform at least 2 db for PSNR and 0.6 on SSIM. On the other hand, SRGAN~\cite{ledig2017photo} which is not optimized for PSNR, mainly focuses on color correction and not much on noise cancellation. Figure~\ref{fig:compare_final} illustrates qualitative comparison of these methods trained on ImagePairs dataset. Needless to say, these models perform much better on nose cancellation, color correction and super resolution compared to similar models trained on DIV2K.

\begin{table}
\begin{center}
\begin{tabular}{l|c|c|c}
\toprule
    Model & Train data & PSNR (db) & SSIM \\
\hline
    Bicubic  & -  & 21.451 & 0.712   \\
    SRGAN~\cite{ledig2017photo}  & DIV2K  & 21.906 & 0.699  \\
    WDSR~\cite{yu2018wide} & DIV2K  &  21.299 & 0.697 \\
    EDSR~\cite{Lim_2017_CVPR_Workshops} & DIV2K  & 21.298 & 0.697  \\
    SRGAN~\cite{ledig2017photo}  & ImagePairs  & 22.161 & 0.673 \\
    WDSR~\cite{yu2018wide} & ImagePairs  & 23.805 & 0.767 \\ 
    EDSR~\cite{Lim_2017_CVPR_Workshops} & ImagePairs  &  23.845 & 0.764 \\
\bottomrule
\end{tabular}
\end{center}
\caption{Comparisons of state-of-the-art single image super resolution algorithms on ImagePairs data set. } 
\vspace{-.25cm}
\label{tab:sr_comapre}
\end{table}

\begin{figure*}
    \vspace{-.40cm}
	\begin{center}
	   \setlength{\tabcolsep}{2pt} % Default value: 6pt
		\begin{tabular}{ccccc}
			Raw & DeepISP~\cite{Schwartz2019DeepISP} & SIDnet~\cite{ChenSid_CVPR2018} & GuidanceNet~\cite{Guidance_Network} & Ground Truth \\
			\includegraphics[width=3.3cm]{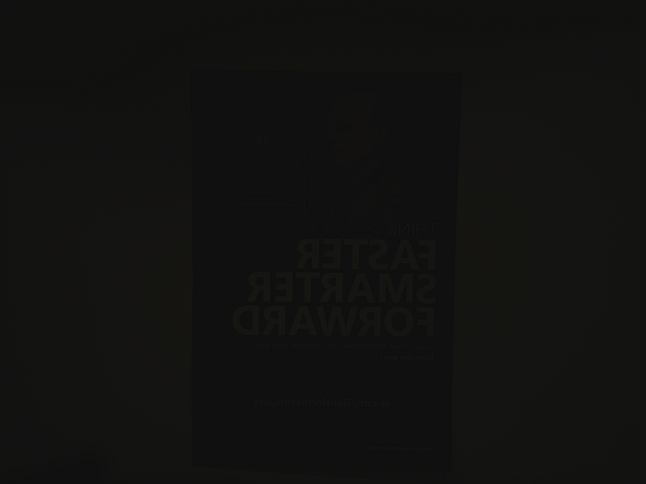} &
			\includegraphics[width=3.3cm]{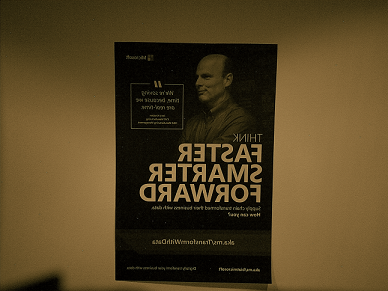} &
			\includegraphics[width=3.3cm]{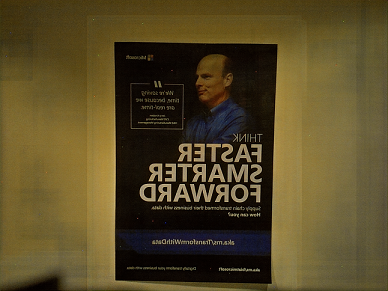} &
			\includegraphics[width=3.3cm]{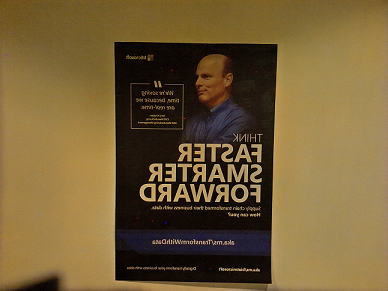} &
			\includegraphics[width=3.3cm]{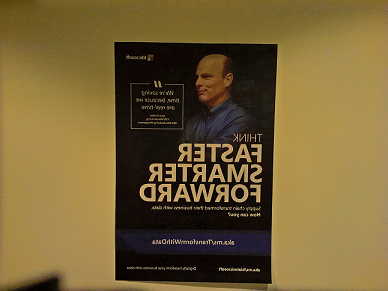} \\
			\includegraphics[width=3.3cm]{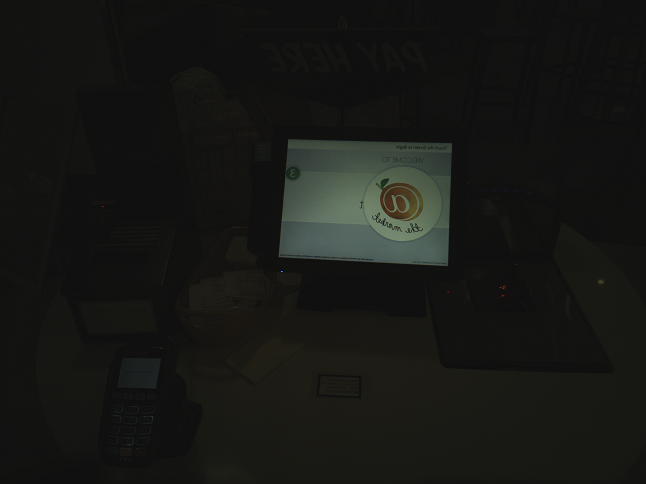} &
			\includegraphics[width=3.3cm]{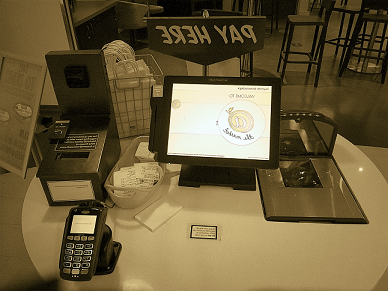} &
			\includegraphics[width=3.3cm]{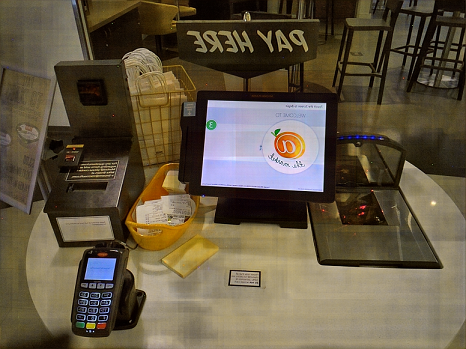} &
			\includegraphics[width=3.3cm]{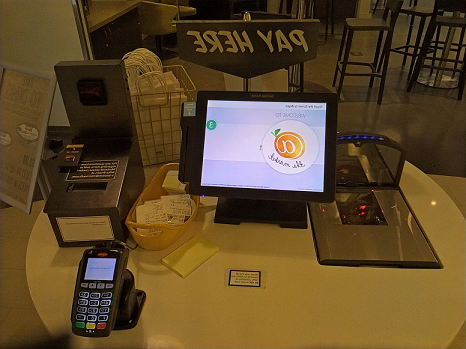} &
			\includegraphics[width=3.3cm]{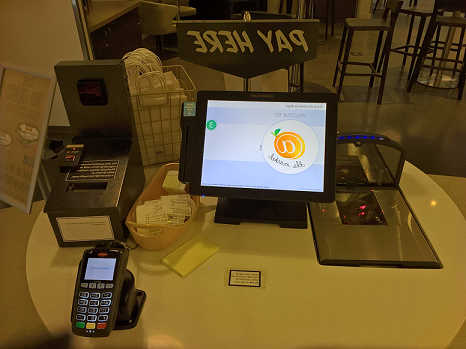} \\
		\end{tabular}
		\caption{Qualitative comparisons of state-of-the-art ISP methods trained on ImagePairs dataset.}
		\label{fig:isp}
	\end{center}
    \vspace{-.25cm}
\end{figure*}

\subsection{ISP Benchmark}
%\vspace{-.25cm}
For ISP task, we consider LR images and their corresponding raw images of ImagePairs train/test sets as the raw HR images are too large. 
We trained DeepISP net~\cite{Schwartz2019DeepISP}, SID net \cite{ChenSid_CVPR2018} and GuidanceNet~\cite{Guidance_Network} on ImagePairs training set which contains raw and LR images. All networks read RAW images and associated 4 camera properties: analogue gain, digital gain, exposure time and lens position. Here, the exposure time is in microsecond; the lens position is the distance between the camera and the scene in centimeters. GuidanceNet~\cite{Guidance_Network} is designed to use camera properties in its bottleneck layers, but we modified DeepISP net \cite{Schwartz2019DeepISP} and SID net \cite{ChenSid_CVPR2018}. For DeepISP net \cite{Schwartz2019DeepISP}, we tile and concatenate these features with the output of their local sub-network and then feed it to the global sub-network for estimating the quadratic transformation coefficients. For SID net \cite{ChenSid_CVPR2018}, we tile and concatenate these features with the input image. Tables~\ref{tab:isp} reports the evaluation of these three models on ImagePairs test set in term of PSNR and SSIM. This shows GuidanceNet~\cite{Guidance_Network} which properly used camera properties outperform others. Figure~\ref{fig:isp} illustrates examples for each of these models.
	
\begin{table}
\begin{center}
\begin{tabular}{l|c|c}
\toprule
    Model & PSNR (db) & SSIM \\
\hline
    DeepISP~\cite{Schwartz2019DeepISP}  & 20.30 &0.89 \\
    SIDnet~\cite{ChenSid_CVPR2018} & 23.08 & 0.90 \\
    GuidanceNet~\cite{Guidance_Network} & 29.22 & 0.96 \\
\bottomrule
\end{tabular}
\end{center}
\caption{Comparisons of ISP algorithms on ImagePairs dataset. } 
\vspace{-.25cm}
\label{tab:isp}
\end{table}

\section{Conclusion} \label{sec:seven}
	In this paper we proposed a new data acquisition technique which could be used as an input for SR, noise cancellation and quality enhancement techniques. We used a beam-splitter to capture the same scene by a low resolution camera and a high resolution camera. Unlike current small-scale datasets used for these tasks, our proposed dataset includes 11,421 pairs of low-resolution and high-resolution images of diverse scenes. Since we also release the raw images, this large-scale dataset could be used for other tasks such as ISP generation. We trained state-of-the art methods for SR and ISP tasks on this dataset and showed how the new dataset can be successfully used to improve the quality of real-world image super resolution significantly.
	
{\small
\bibliographystyle{ieee_fullname}
\bibliography{sr}
}
	
\end{document}